\documentclass[epj,final]{svjour}

\usepackage{graphicx}

\begin{document} 
 
\title{Scaling behavior of a nonlinear oscillator with 
additive noise, white and colored}

\author{Kirone Mallick\inst{1} \and Philippe Marcq\inst{2}}                 

\institute{Service de Physique Th\'eorique, Centre d'\'Etudes de Saclay,
 91191 Gif-sur-Yvette Cedex, France \\
\email{mallick@spht.saclay.cea.fr}\and 
Institut de Recherche sur les Ph\'enom\`enes Hors \'Equilibre,
Universit\'e de Provence,\\
49 rue Joliot-Curie, BP 146, 13384 Marseille Cedex 13, France\\
 \email{marcq@irphe.univ-mrs.fr}}

\date{December 13, 2002}

\abstract{
We study analytically and numerically the problem of a nonlinear mechanical
oscillator with additive noise in the absence of  damping. 
We show that the amplitude, the velocity  and  the energy  
of the oscillator  grow  algebraically with time. 
For Gaussian  white noise,  an  analytical  expression for the
probability distribution function of the energy is obtained in the 
long-time limit. In the case of colored, Ornstein-Uhlenbeck noise,  
a self-consistent  calculation leads to (different) anomalous 
diffusion exponents.
Dimensional analysis yields the qualitative behavior of 
the prefactors (generalized diffusion constants)
 as a function of the  correlation time.
} 

\maketitle 

 \section{Introduction}
\label{sec:intro}

The concept of noise in Physics stems from the  theory of Brownian motion 
in which the force exerted on a macroscopic impurity  by  the surrounding 
molecules is represented by a random function ${\cal F}(t)$ \cite{wax}.
This model provides a microscopic explanation  of the law  of  diffusion 
and yields  the  celebrated  fluctuation-dissipation  relation 
\cite{vankampen}.
Assuming that the Brownian particle of position $x(t)$ and velocity $v(t)$
is also submitted to a harmonic potential 
${\cal U}(x) = \frac{1}{2} \, \omega^2 \, x^2$, 
the dynamic equation associated with this system  is 
the linear Langevin equation:
\begin{equation}
   \; \frac{\textrm{d} }{\textrm{d} t}v(t) 
   + \gamma \,  v(t) + \omega^2 \, x(t)  = {\cal F}(t) \,,
 \label{Langevin}
\end{equation}
where $\gamma$ is the effective damping coefficient.

The interplay between  noise and  nonlinearity  genera\-tes  
many original  phenomena  that make the study of  random
dyna\-mical systems a  rich and  fascinating field.
Noise in a nonlinear system  may  induce non-equilibrium
phase transitions \cite{lefever,landaMc}
or improve the performance of a device via stochastic 
resonance \cite{marchesoni}.  
Random fluctuations may be rectified into a directed motion
when a particle in a fluctuating environment is submitted
to a ``ratchet-like'' potential  \cite{reimann}.  

  A  simple  nonlinear noisy dynamical system  is obtained by 
   considering  Eq.~(\ref{Langevin}) with  an anharmonic
potential ${\cal U}(x)$. If the random force is  approximated
by  a Gaussian white noise, the stationary probability distribution function 
(PDF) of the energy,
valid when $t \gg \gamma^{-1}$, is given by the
  cano\-nical  distribution.
To the best of our knowledge, the closed form of the distribution function 
of the nonlinear oscillator's energy is not known when the 
correlation time $\tau$ of 
the random force is non-zero (see \cite{colored} and references therein
for approximate results valid in the small $\tau$ limit).

In this work, we study the  motion of an \emph{undamped}  
nonlinear oscillator submitted to an additive noise.
For Gaussian white noise (Section \ref{sec:white}), 
we apply the method presented in a recent article \cite{philkir},
where we studied the dyna\-mical behavior of an undamped nonlinear oscillator 
with a fluctuating frequency represented as a  parametric noise.
We calculate  the  probability distribution function of the energy
in the long-time limit 
and match it to the  distribution obtained in presence 
of damping. The average energy, root-mean-square amplitude and velocity
of the oscillator grow algebraically with time, with anomalous diffusion 
exponents different from those obtained for parametric noise.
We show in Section \ref{sec:colored} that diffusion
is reduced  at large times when  the additive noise has a non-zero
correlation time. Anomalous
diffusion exponents are calculated in a self-consistent manner
for an Ornstein-Uhlenbeck random force.
In a somewhat counter-intuitive fashion, the white-noise behavior 
is  observed for sufficiently \emph{short} times.
The crossover from white to co\-lored noise
can be embodied in a single scaling function  as shown by
 dimensional analysis.
All  the results obtained  for a nonlinear oscillator submitted to additive
or multiplicative, white or colored Gaussian noise are summarized in 
Table 1.

 \section{The nonlinear oscillator with additive white noise}
 \label{sec:white}

  We study  a particle trapped in a confining potential ${\cal U}(x)$
 and subject to additive noise. We neglect dissipative effects unless
 stated otherwise.   The dynamics of the nonlinear oscillator is  given by
 \begin{equation}
   \frac{\textrm{d}^2 }{\textrm{d} t^2}x(t) 
   = - \frac { \partial{\mathcal U}(x)}{\partial x}  + \xi(t) \,,
 \label{dynwhite}
\end{equation}
where $x(t)$ represents the position of the oscillator at time $t$.
In this Section, the random noise  $\xi(t)$  is a Gaussian white noise 
 of zero mean-value and of amplitude  ${\mathcal D}$:
 \begin{eqnarray}
       \langle \xi(t)  \rangle &=&   0   \, ,\nonumber \\
   \langle \xi(t) \xi(t') \rangle  &=&  {\mathcal D} \, \delta( t - t') .
   \label{defgamma}
 \end{eqnarray} 

For analytical as well as numerical calculations, we shall interpret 
stochastic differential equations such as (\ref{dynwhite}) according 
to Stratonovich's convention.
Indeed, Strato\-no\-vich  calculus appears naturally 
in the limit of a vani\-shingly  small correlation time \cite{vankampen}.
It must be emphasized that this convention (as opposed to
Ito calculus) introduces correlations between the noise 
and the dynamical  variables at the same time $t$ \cite{itostrato}. 
 Moreover, the  mechanical oscillator defined by Equations 
(\ref{dynwhite},\ref{defgamma}) is \emph{not} conservative:
noise feeds energy into the system (see Eqs.~(\ref{evolnE})
and (\ref{moyE})).
 In the absence of  damping, the oscillator's
  amplitude increases  with time.

 For the potential ${\cal U}(x)$ to be confining, we must have 
  ${\mathcal U} \rightarrow +\infty$ when $ |x| \rightarrow \infty$.   
 We shall restrict our analysis to the case  where ${\mathcal U}(x)$
 is a polynomial, and in order to respect the $ x \to -x$
symmetry, ${\cal U}(x)$ is taken to be  even in $x$.
 Hence, when  $ |x| \rightarrow \infty$,
\begin{equation}
     {\mathcal U}   \sim \frac{ x^{2n}}{2n}
  \,\, \hbox{ with } \,\, n \ge  2 \,,   
 \label{infU}
\end{equation}
 where  the coefficient of $x^{2n}$  has been set to  $1/(2n)$
 after  a suitable rescaling.
  As the amplitude $x$  of the oscillator
 grows at large times, only the asymptotic behavior 
 of ${\mathcal U}(x)$ for $ |x| \rightarrow \infty$ 
 is  relevant  and thus 
 Eq.~(\ref{dynwhite}) reduces to  
  \begin{equation}
   \frac{\textrm{d}^2 }{\textrm{d} t^2}x(t)  + x(t)^{2n-1}  = \xi(t)   \,.
 \label{nthorder}
\end{equation}  
 
The  method  we shall use to study 
Eqs.~(\ref{defgamma},\ref{nthorder}) is akin to that
presented  in our previous work on nonlinear oscillators with
multiplicative noise  \cite{philkir}.  
In Section \ref{sec:white:aa}, we use the integrability properties 
of the deterministic nonlinear
oscillator to write exact stochastic differential
equations in energy-angle variables.  
We then  derive equipartition relations in Section 
\ref{sec:white:equip}. Averaging out the fast angular variable
\cite{strato,lindenberg}, we calculate the energy's probability distribution
function in the long-time limit (Section \ref{sec:white:pdf}). 
In Section \ref{sec:white:scaling}, we obtain the associated 
anomalous diffusion exponents and constants and study their dependence
 on the stiffness of the confining potential at infinity.
Our numerical simulations are based on a time
discretization presented in \cite{mannella}, and described in detail
in \cite{philkir}.

 \subsection{Energy-angle coordinates}
\label{sec:white:aa}

We rewrite the second-order stochastic  differential
 equation (\ref{nthorder})   as a first order 
system in energy (or action)  and angle  variables.
Without noise, the deterministic version of Eq.~(\ref{nthorder}):
\begin{equation}
   \ddot x + x^{2n-1} = 0 \, ,
 \label{ndeter}
\end{equation}
 is integrable because of energy conservation, where the energy 
is defined by
 \begin{equation}
   E =  \frac{1}{2}\dot x^2 + \frac{1}{2n} x^{2n} \, .
\label{energyn}
\end{equation}
 The action-angle variables $(I,\phi)$ associated 
with  Eq.~(\ref{ndeter}) are 
\begin{eqnarray}
        I &=& 4 \int_0^{(2nE)^{\frac{1}{2n}}} 
\sqrt{ 2E - \frac{x^{2n}}{n}} {\textrm d}x  \nonumber \\
  &=& 4 \, (2^{n+1}n)^{\frac{1}{2n}} \, E^{ \frac{n+1}{2n} }
 \int_0^1  \sqrt{ 1 - u^{2n}} {\textrm d}u \nonumber \\
  &\propto& E^{ \frac{n+1}{2n} } \,   ,  \\
      \phi &=&   \sqrt{n} \, \int_0 ^{ x/(2nE)^{{1}/{2n}} }
    \frac{{\textrm d}u}{\sqrt{ 1 - u^{2n}}} \, ,
\end{eqnarray}
 where the  angle  variable  $\phi$  is 
 defined modulo the  oscillation  period  $4 K_n$, with  
\begin{equation}
   K_n   = \sqrt{n} \int_0^1  \frac{{\textrm d}u}{\sqrt{ 1 - u^{2n}}} \, .
\label{nperiod}
\end{equation}

 The solution of Eq.~(\ref{ndeter}) can be parametrized as a function
 of energy $E$ and angle $\phi$
\begin{eqnarray}
          x &=&   E^{1/{2n}} \, {\mathcal S}_n (\phi) ,
  \label{solnx}  \\                 
     \dot x &=& (2n)^{\frac{n-1}{2n}} E^{1/2} \,
  {\mathcal S}_n' ( \phi ) \, ,
\label{solnv}
 \end{eqnarray}
where the hyper-elliptic function ${\mathcal S}_n$ is defined as
   \begin{eqnarray}
{\mathcal S}_n(\phi) = Y \,  \leftrightarrow 
  \phi  &=&  \sqrt{n} \int_0^{ \frac{Y}{(2n)^{1/2n}}} 
 \frac{{\textrm d}u}{\sqrt{ 1 - u^{2n}}}   \nonumber \\
  &=&  \frac{ \sqrt{n}} { (2n)^{1/2n} } \int_0^Y 
   \frac{{\textrm d}u}{\sqrt{ 1 -  \frac{u^{2n}}{2n}}}      \,    .
  \label{hyperelli}
\end{eqnarray}
 From this definition, we obtain  the following relation
 between   ${\mathcal S}_n$ and its derivative ${\mathcal S}_n'$  
\begin{equation}
 {\mathcal S}_n'(\phi) = \frac{ (2n)^{ \frac{1}{2n}}}{\sqrt{n}} \left(
      1 -  \frac{({\mathcal S}_n(\phi))^{2n}}{2n} \right)^{\frac{1}{2}} .
\label{derivS}
\end{equation}

 Writing  Eq.~(\ref{nthorder}) in terms
 of energy and angle variables, we obtain the following
 stochastic dynamical system
\begin{eqnarray}
\dot{E}   &=&   \dot{x} \; \xi(t) = 
 (2n)^{\frac{n-1}{2n}} \, E^{\frac{1}{2} } \, {\mathcal S}_n'(\phi) \, \xi(t) 
\label{evolnE}   \, ,  \\ 
   \dot\phi &=& (2nE)^{\frac{n-1}{2n}} - \frac{1}{ (2n)^{\frac{1}{2n}}} \,
   \frac{  {\mathcal S}_n(\phi)} { (2nE)^{\frac{1}{2}} } \, \xi(t)
\label{evolnphi}  \, .
   \end{eqnarray}
  With the help of  the  auxiliary variable $\Omega$
 \begin{equation}
 \Omega =   (2n)^{ \frac{n+1}{2n} } \,  E^{\frac{1}{2}}  \, ,
 \label{defOmega}
  \end{equation}
 Eqs.~(\ref{evolnE}) and (\ref{evolnphi}) are  written 
   in  the simpler form:
   \begin{eqnarray}
     \dot \Omega  &=& n  \, {\mathcal S}_n'(\phi)  \, \xi(t)   \label{evoln1}
   \,  ,   \\   
 \dot\phi  &=& \Big ( \frac{\Omega}{ (2n)^{\frac{1}{2n}}} 
\Big )^{\frac{n-1}{n}}
 - \frac{{\mathcal S}_n(\phi)}{\Omega}  \, \xi(t)  \, .
    \label{evoln}
   \end{eqnarray}
 These equations have been derived without any hypothesis
 on the function $\xi$ and   are rigorously equivalent
 to the original equation (\ref{nthorder}).

 Although the method is identical, this set of equations differs 
from that obtained for parametric noise (See Eqs.~(27)-(28) in 
\cite{philkir}). As a consequence, additive and multiplicative
random noises
lead to different long-time  behaviors of the observables of the system
(See table 1).

\begin{figure}
\includegraphics*[width=0.75\columnwidth]{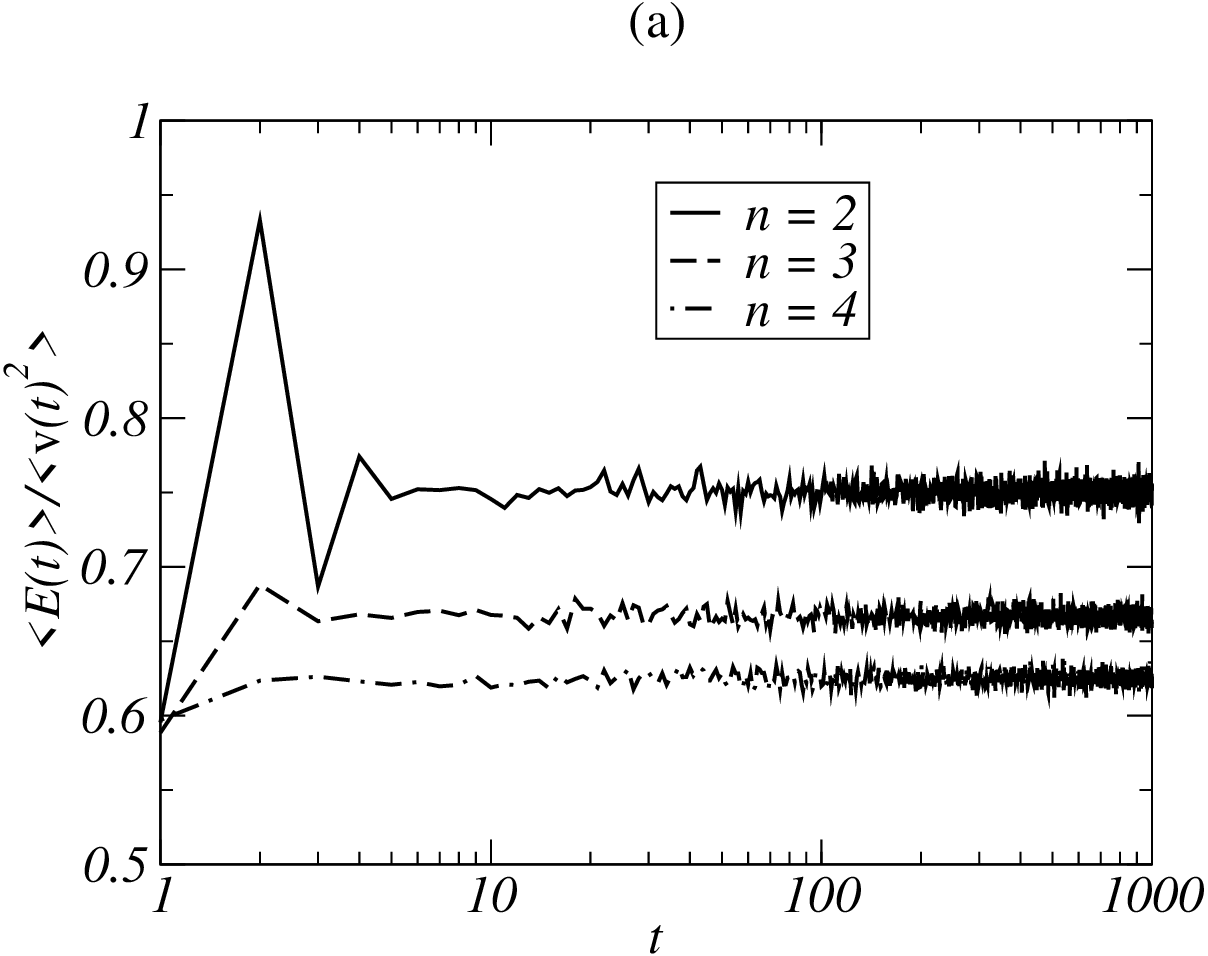}
\bigskip

\includegraphics*[width=0.75\columnwidth]{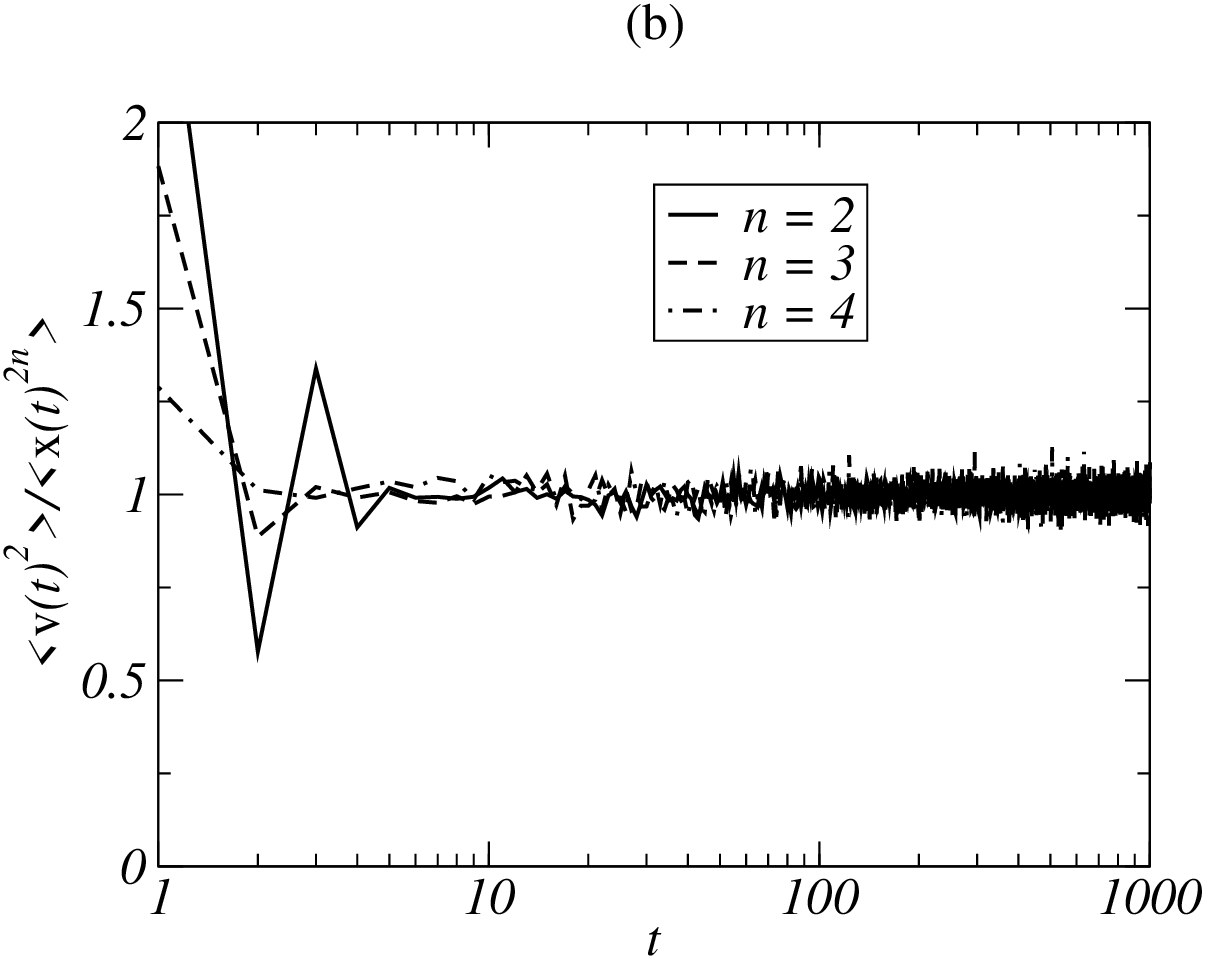}

\caption{Nonlinear oscillator with additive white noise:
Eq.~(\ref{nthorder}) is integrated numerically for ${\mathcal D} = 1$, 
with a timestep $\delta t$, and averaged over $10^4$ realizations for
$n=2$ ($2n - 1 = 3$), $\delta t = 5 \, 10^{-4}$;
$n=3$ ($2n - 1 = 5$), $\delta t = 5 \, 10^{-4}$;
$n=4$ ($2n - 1 = 7$), $\delta t = 2 \, 10^{-4}$.
Fig.~(a): the first equipartition ratio 
$\langle E(t) \rangle / \langle v(t)^2 \rangle$ is close to
the theoretical value $\frac{n+1}{2n}$  given in  Eq.~(\ref{equipn2}):
$3/4$ for $n=2$; $2/3$ for $n=3$; $5/8$ for $n=4$.
Fig.~(b): the second equipartition ratio 
$\langle v(t)^2 \rangle / \langle x(t)^{2 n} \rangle$ is close to $1$
for $n = 2,3,4$.
}
\label{fig:white:equip} 
\end{figure}

\subsection{Equipartition relations}
\label{sec:white:equip}

Noting that ${\mathcal S}_n$ and ${\mathcal S}'_n$ are bounded 
functions of $\phi$,  we deduce from Eqs.~(\ref{evoln1}) 
and (\ref{evoln}) that
$\Omega$ is a diffusive variable that scales as  $t^{1/2}$,
 whereas $\phi$ is a fast variable that scales like $t^{3/2 -1/2n}$ 
 (these assertions will be justified  rigorously in the next  section).
 In the long time limit, it is therefore justified to average
 the dynamics  over  the variations of the fast variable $\phi$.
Using Eqs.~(\ref{solnx})-(\ref{solnv}), and assuming that 
$\phi$ is uniformly distributed over the 
 interval $[0, 4K_n]$ of a period, we find the following relations:
\begin{eqnarray}
\langle  E   \rangle   &=&  \frac{n+1}{2n} \, \langle \dot x^2   \rangle \, ,
\label{equipn2}  \\
\langle \dot x^2   \rangle   &=&   \langle  x^{2n}   \rangle  \, .
\label{equipnxv}
\end{eqnarray}
Figure~\ref{fig:white:equip} shows that Eqs.~(\ref{equipn2})-(\ref{equipnxv})
are indeed verified with excellent accuracy.

Because the transformation of variables (\ref{solnx})-(\ref{solnv})
is independent of the nature of the noise $\xi(t)$, 
the statistical identities (\ref{equipn2})-(\ref{equipnxv}) 
 are the same as those obtained for parametric noise \cite{philkir}.
Such generalized equipartition relations are valid whenever the asymptotic
probability distribution function is independent  of the angle variable
and is a function of the energy only. This is the case, in particular,
of the canonical     Boltzmann-Gibbs    distribution: 
Eqs.~(\ref{equipn2})-(\ref{equipnxv}) are valid at 
thermodynamic equilibrium, as can be shown 
 using standard techniques of statistical mechanics 
in the canonical ensemble.

\begin{figure}
\includegraphics*[width=0.75\columnwidth]{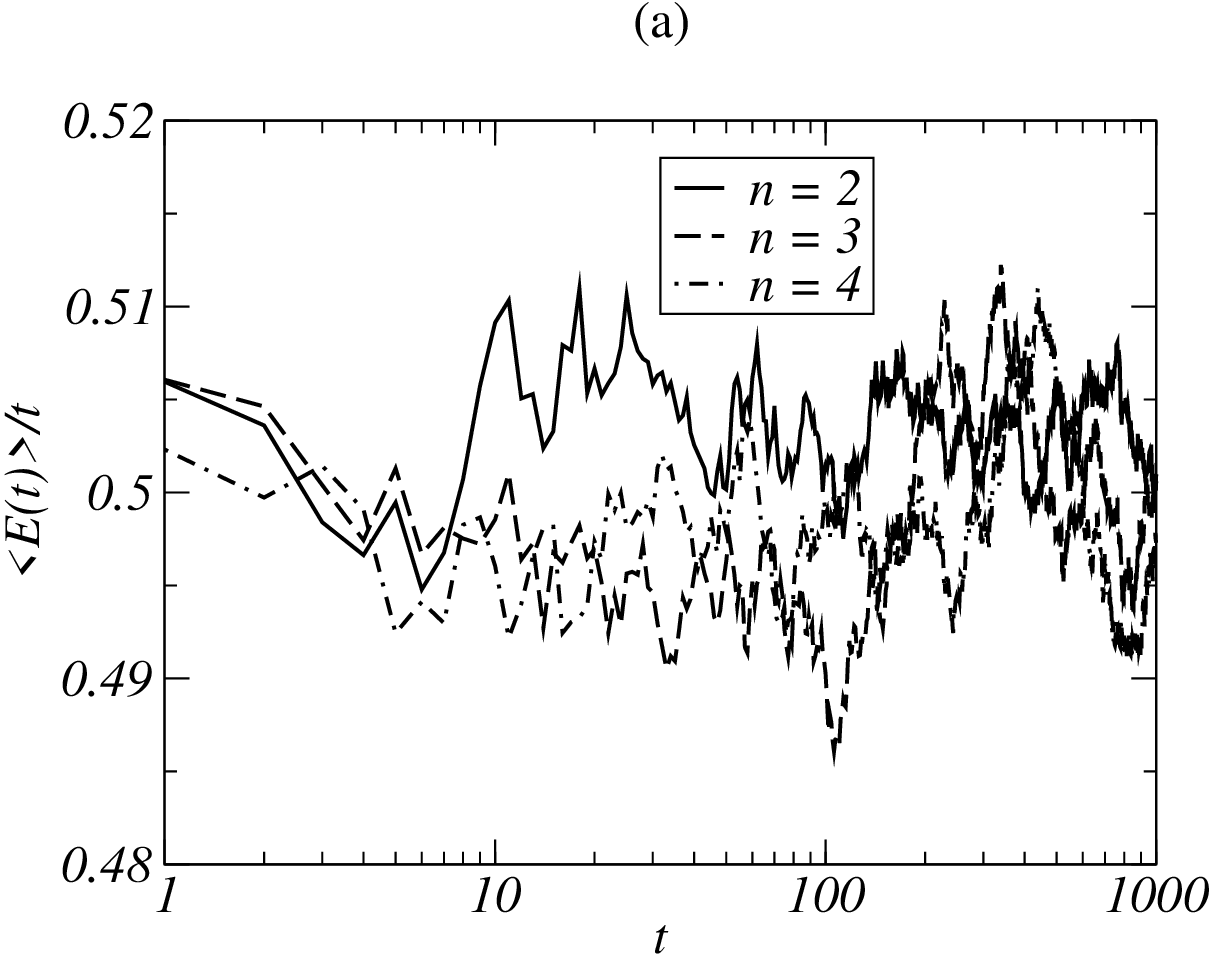}
\bigskip

\includegraphics*[width=0.75\columnwidth]{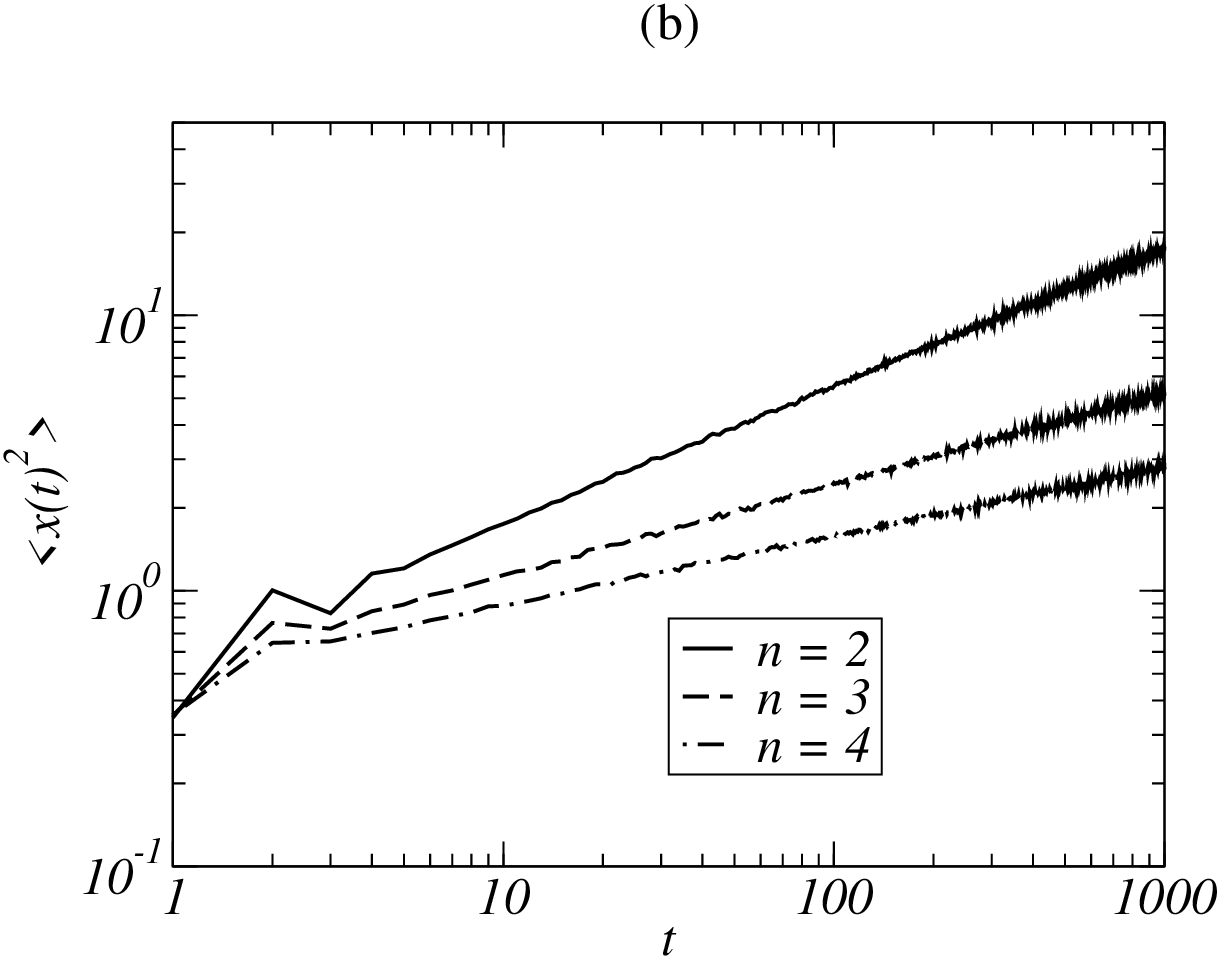}
\bigskip

\includegraphics*[width=0.75\columnwidth]{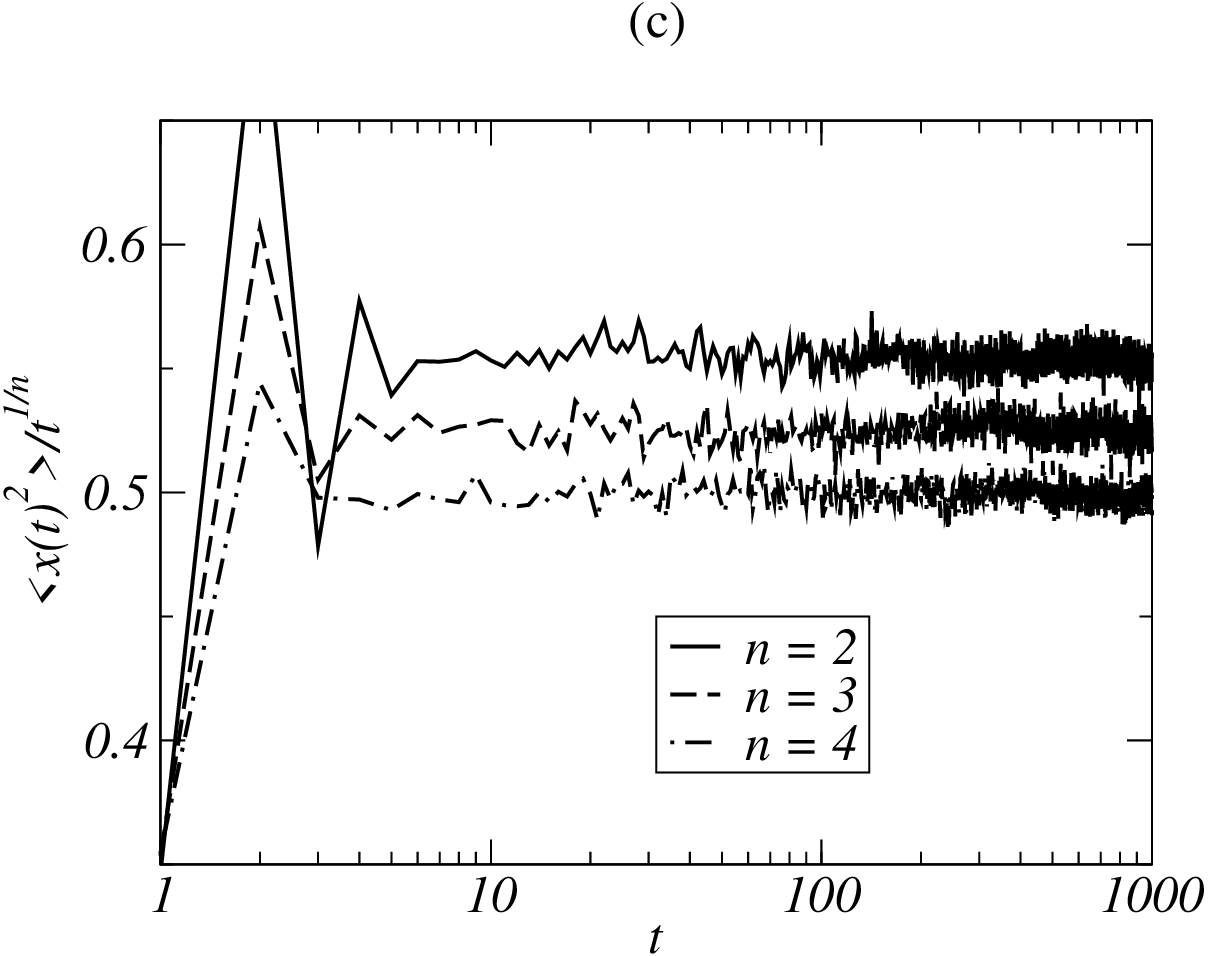}

\caption{Nonlinear oscillator with additive, Gaussian white noise.
Fig.~(a): the ratio $\langle E \rangle/t$ is found close to ${\cal D}/2$
(${\cal D} = 1$) for $n = 2,3,4$.
Fig.~(b): scaling behavior of $\langle x(t)^2 \rangle$.
Fig.~(c):  the ratio $\langle x^2 \rangle/t^{1/n}$ yields the 
following generalized diffusion constants
$D_x^{(n)}$: $D_x^{(2)} = 0.553(5)$, 
$D_x^{(3)} = 0.525(5)$, $D_x^{(4)} = 0.499(4)$,
in excellent agreement with the predictions of
 Eqs.~(\ref{quartic})-(\ref{hepta}).
Numerical data is obtained from the same simulations as in 
Fig.~\ref{fig:white:equip}.
}
\label{fig:white} 
\end{figure}

\subsection{Calculation of the  effective PDF}
\label{sec:white:pdf}

 The  Fokker-Planck equation corresponding to the system (\ref{evoln1},
\ref{evoln})  reads
 \begin{eqnarray}
  \partial_t &P& =
 -\Bigg ( \frac{\Omega}{ (2n)^{\frac{1}{2n}}} \Bigg )^{(n-1)/n} 
  \partial_{\phi}P \label{FPomeg}  \\
 &+&   \frac{{\mathcal D}}{2}
\Bigg( \partial_{\phi}\Big( \frac{g(\phi)}{\Omega}  \partial_{\phi}
  \frac{g(\phi)}{\Omega} P \Big)
 - \partial_{\phi}\Big(  \frac{g(\phi)}{\Omega} 
   \partial_{\Omega}   f(\phi) P     \Big)  \nonumber\\
 &&- \partial_{\Omega}\Big( f(\phi) 
 \partial_{\phi} \frac{g(\phi)}{\Omega}P \Big)  
+ \partial_{\Omega}\Big( f(\phi) \partial_{\Omega}f(\phi)P    \Big)  
\Bigg)   \nonumber
 \end{eqnarray}
where we have defined
 \begin{equation}
  f(\phi)  =  n  \, {\mathcal S}_n'(\phi) 
         \,\,\, \hbox { and }  \,\,\, 
  g(\phi) = {\mathcal S}_n(\phi)   \,\, .
\end{equation}  
This equation  governs the  dynamics of the  probability 
distribution function $P_t(\Omega, \phi)$. As we already noticed, 
  the angle variable $\phi$ varies rapidly
  as  $\Omega$ grows.   
We carry out  the averaging of Eq.~(\ref{FPomeg}) 
under the hypothesis that  the probability density,  $P_t(\Omega, \phi)$, 
becomes uniform in $\phi$  when $ t \to \infty$. 
The averaged evolution
equation of the reduced probability density $\tilde{P}_t(\Omega)$ is then
 \begin{equation}
   \partial_t \tilde{P}  =  \overline{f^2(\phi)}\;  \frac{ {\mathcal D} }{2} 
\left(  \partial_{\Omega}^2 \tilde{P} - \frac{1}{n} \partial_{\Omega}
      \frac{\tilde{P}}{\Omega}  \right)  \, ,
\label{FPmoy}
\end{equation}  
where we have used   $f  =  (\partial_{\phi} g)/n $. The
  notation $\overline{f^2(\phi)}$ re\-presents the mean value
 of the function $f^2$ over one period $4K_n$ of $\phi$.
 This constant has been calculated explicitly in \cite{philkir}
 and   is given by
\begin{equation}
 \overline{f^2(\phi)} = n^2 \; \overline{{\mathcal S}_n'(\phi)^2} = 
 \frac{ n^2  \, (2n)^{\frac{1}{n}} }{ n + 1 }  \, .
\end{equation}  

The Fokker-Planck equation (\ref{FPmoy}) is equivalent 
to  an effective  first-order stochastic differential equation 
 for the slow variable $\Omega$.
 Defining  an effective Gaussian white noise $\tilde{\xi}(t)$
 such that
\begin{equation}
 \langle \tilde\xi(t) \tilde\xi(t') \rangle  =  
   \tilde{{\mathcal D}} \, \delta( t - t')  \,\,\,\, \hbox{ with } \,\,\,\, 
  \tilde{{\mathcal  D}} =  \frac{ n^2  (2n)^{\frac{1}{n}} }{ n + 1 }
 {\mathcal  D}  \, ,
\label{defDtilde}
\end{equation}  
 we deduce from Eq.~(\ref{FPmoy}) the effective Langevin equation for $\Omega$
\begin{equation}
 \dot \Omega = \frac{  \tilde{{\mathcal D}} }{2 \; n}
  \, \frac{1}{\Omega}  + \tilde\xi(t)  \, .
\label{nLangeff} 
\end{equation}  
Thus $\Omega$ can be reinterpreted  as a Brownian variable in a logarithmic
 potential. 

The scale-invariant effective Fokker-Planck equation (\ref{FPmoy}) 
can be solved explicitly. We find
\begin{equation}
   {\tilde P}_t(\Omega) =
  \frac{2}{  \Gamma \left(\frac{n + 1}{2n}\right)} \,
 \frac{ \Omega^{\frac{1}{n}}  \,  
{\rm e}^{ -  \Omega^2/(2\tilde{{\mathcal D}} t) }  }
{ (2\tilde{{\mathcal D}} t )^{\frac{n  + 1}{2n}}    }    \, ,
\label{pdf2Om}
\end{equation}
where $\Gamma$ is the Euler Gamma function.
 Using  Eq.~(\ref{defOmega}), we deduce  from Eq.~(\ref{pdf2Om}) the 
 asymptotic  probability distribution function  of the
 energy variable:
\begin{equation}
\label{pdf2En}
   {\tilde P}_t(E) =
  \frac{1}{  \Gamma \left(\frac{n + 1}{2n}\right)} \,
 \frac{1}{ E } \, \Big( \frac{  (2n)^{\frac{n  + 1}{n}} E}
  { 2\tilde{{\mathcal D}} t } \Big)^{\frac{n  + 1}{2n}}  
   \exp\left\{ 
- \frac{ (2n)^{\frac{n  + 1}{n}} E }{2\tilde{{\mathcal D}} t} \right\}
\end{equation}

 Moreover, the P.D.F. (\ref{pdf2En}) can be matched to the cano\-nical 
 distribution. Suppose that some damping is pre\-sent in the system.
 The dynamical equation (\ref{nthorder}) now reads
 \begin{equation}
   \frac{\textrm{d}^2 }{\textrm{d} t^2}x(t)  + 
 \gamma  \frac{\textrm{d} }{\textrm{d} t}x(t)
  + x(t)^{2n-1}  = \xi(t)   \,,
 \label{nthorderdamp}
\end{equation}  
 where $\gamma$ is the friction coefficient. The stationary
 distribution associated with  the stochastic equation (\ref{nthorderdamp})
 is:
\begin{equation}
  {\tilde P}_{\textrm{can}}(E) = 
 \frac{1}{  \Gamma \left(\frac{n + 1}{2n}\right)} \, \frac{1}{ E } \,
  \Big( \frac{ 2 \gamma E }{{\mathcal D}} \Big)^{\frac{n  + 1}{2n}}  
 \exp \left\{ -  \frac{2 \gamma E}{  {\mathcal D}}  \right\}  \,.
 \label{Gibbs}
\end{equation}   
This PDF  becomes the Boltzmann-Gibbs  cano\-nical  measure
 when supplemented with the fluctuation-dissipation relation \cite{vankampen}.
The crossover time  $t_c$ from  the  asymptotic distribution function
 for the undamped oscillator to the Gibbs  measure
 is found by matching Eq.~(\ref{pdf2En}) with  Eq.~(\ref{Gibbs}):
 \begin{equation}
 t_c  = \frac{n+1}{2n \gamma}\, .
\end{equation}  
 Thus, there are three distinct time scales involved: a fast time scale
 over which the angle variable $\phi$ gets uniformly distributed,
 a `long'  time scale  over which the effective Fokker-Planck
 equation for $\Omega$ (\ref{FPmoy}) is relevant, and finally times
 longer than the crossover time  $t_c$  beyond which the
 inevitable damping in the system takes over.

\subsection{Time-asymptotic behavior of observables}
\label{sec:white:scaling}

  With  the   probability  distribution function  
  (\ref{pdf2En}), we can 
 calculate the long time behavior, in the absence of damping,
  of the expectation value 
 of  any observable that depends on the 
 position and velocity  of the particule.  In particular, we obtain
\begin{eqnarray}
  \langle x^2  \rangle  &=&   
 \frac{ \Gamma\Big( \frac{3}{2n} \Big) }
       { \Gamma\Big( \frac{1}{2n} \Big)}
    \left( \frac{ 2 \; n^2 }{n+1} \; {\mathcal D}  t \right)^ {\frac{1}{n}} 
 \label{moyx2} \, ,\\
\langle \dot{x}^2  \rangle  &=&  \frac{n {\mathcal D} }{n+1} t 
 \label{moyv2} \,, \\
 \langle E  \rangle  &=&  \frac{ {\mathcal D} }{2} t 
 \label{moyE} \,.
\end{eqnarray}
 From Eq.~(\ref{moyx2}), we notice  that 
the particle diffuses more slowly in a stiffer potential well,
as one would intuitively expect. 
The  equipartition relation  between energy and velocity obtained in 
Section~\ref{sec:white:equip} is confirmed by
Eqs.~(\ref{moyv2},\ref{moyE}).
 For \emph{all} potentials growing algebraically at infinity,
we find that the mean value of  the energy grows linearly with time, 
with a diffusion constant equal to  ${\mathcal D} /2 $
(Eq.~(\ref{moyE})). Such a  remarkable and  simple behavior  was not obvious
 a priori. 

 These predictions are confirmed by numerical simulations.
 The universal behavior of  $\langle E  \rangle$ is represented 
in Fi--gure~\ref{fig:white}.a. 
 Using a numerical value of the noise amplitude ${\mathcal D} = 1$
  in Eq.~(\ref{moyx2}), we obtain 
 \begin{eqnarray}
 &\hbox{For $ n =2  $},& \,\,\, 
  \langle  {x}^2  \rangle  =   0.552  \;t^{1/2}  \,\,\,  ,
 \label{quartic} \\
&\hbox{For $ n =3 $},& 
  \,\,\,  \langle  {x}^2  \rangle  =   0.526  \;t^{1/3}      \,\,\,  , 
 \label{penta}   \\
  &\hbox{For $ n =4 $},& 
   \,\,\,  \langle x^2  \rangle  =   0.500  \; t^{1/4}      \,\,\,  . 
\label{hepta}
\end{eqnarray}
These formulae  are in excellent agreement with the numerical results
 displayed in Figures~\ref{fig:white}.b and \ref{fig:white}.c. 

 The validity of  Eqs.~(\ref{moyx2},\ref{moyv2},\ref{moyE}) 
 can be checked in the linear case,  $n=1$.
The exact solution of the equation
\begin{equation}
   \frac{\textrm{d}^2 }{\textrm{d} t^2}x(t)  + \omega^2 x(t)  = \xi(t)   \,,
 \label{linear}
\end{equation}  
 is given by
\begin{eqnarray}
     x(t)  &=&  \frac{1}{\omega} 
 \int_0^t  \sin \left(\omega( t -u)\right) \; \xi(u) \; \rm{d}u\,,
 \label{linx}\\
    \dot{x}(t)  &=& \int_0^t \cos\left( \omega( t -u)\right) \; \xi(u)
\; \rm{d}u  \, .
 \label{linv}
\end{eqnarray}
We obtain the following  exact formulae  \cite{vankampen}:
\begin{eqnarray}
\langle x^2  \rangle  &=&   \frac{ {\mathcal D} }{ 2 \omega^2} t\;
 \left( 1 - \frac{   \sin 2\omega t }{   2\omega t }  \right)   \,,     \\
 \langle \dot{x}^2  \rangle  &=&  \frac{ {\mathcal D} }{ 2 }  t \;
 \left( 1 + \frac{   \sin 2\omega t }{   2\omega t }  \right)    \, ,  \\
 \langle E  \rangle  &=&  \frac{ {\mathcal D} }{ 2 } t \, .
\end{eqnarray}
In the long time limit  $t \rightarrow \infty$,
these expressions agree with Eqs.~(\ref{moyx2},\ref{moyE})
when one substitutes $n =1$ (and $\omega =1$).

 \section{The nonlinear oscillator with additive colored  noise}
\label{sec:colored}

   In order to model an additive  Gaussian noise 
with a non vanishing correlation time $\tau$, we   use
an  Ornstein-Uhlenbeck process, denoted by $\eta(t)$. 
We thus  replace Eq.~(\ref{nthorder})  by the following equation
\begin{equation}
   \frac{\textrm{d}^2 }{\textrm{d} t^2}x(t)  + x(t)^{2n-1}  = \eta(t)   \,.
 \label{dyncolor}
\end{equation}
where the random   noise  $\eta(t)$  is obtained from
 \begin{equation} 
 \frac{{\textrm d} \eta(t)}{{\textrm d} t} = -\frac{1}{\tau} \eta(t) + 
\frac{1}{\tau} \xi(t) \, ,
  \label{OU}
\end{equation} 
and $\xi(t)$ is the Gaussian white noise defined in Eq.~(\ref{defgamma}).
  The stationary  statistical properties of  $\eta$  are given  by
 \begin{eqnarray}
       \langle \eta(t)  \rangle &=&   0 \, ,\nonumber \\
   \langle \eta(t) \eta(t') \rangle  &=&   
\frac{\mathcal D}{2 \, \tau}   \, {\rm e}^{-|t - t'|/\tau} \,.
   \label{deftau}
 \end{eqnarray} 

\begin{figure}
\includegraphics*[width=0.75\columnwidth]{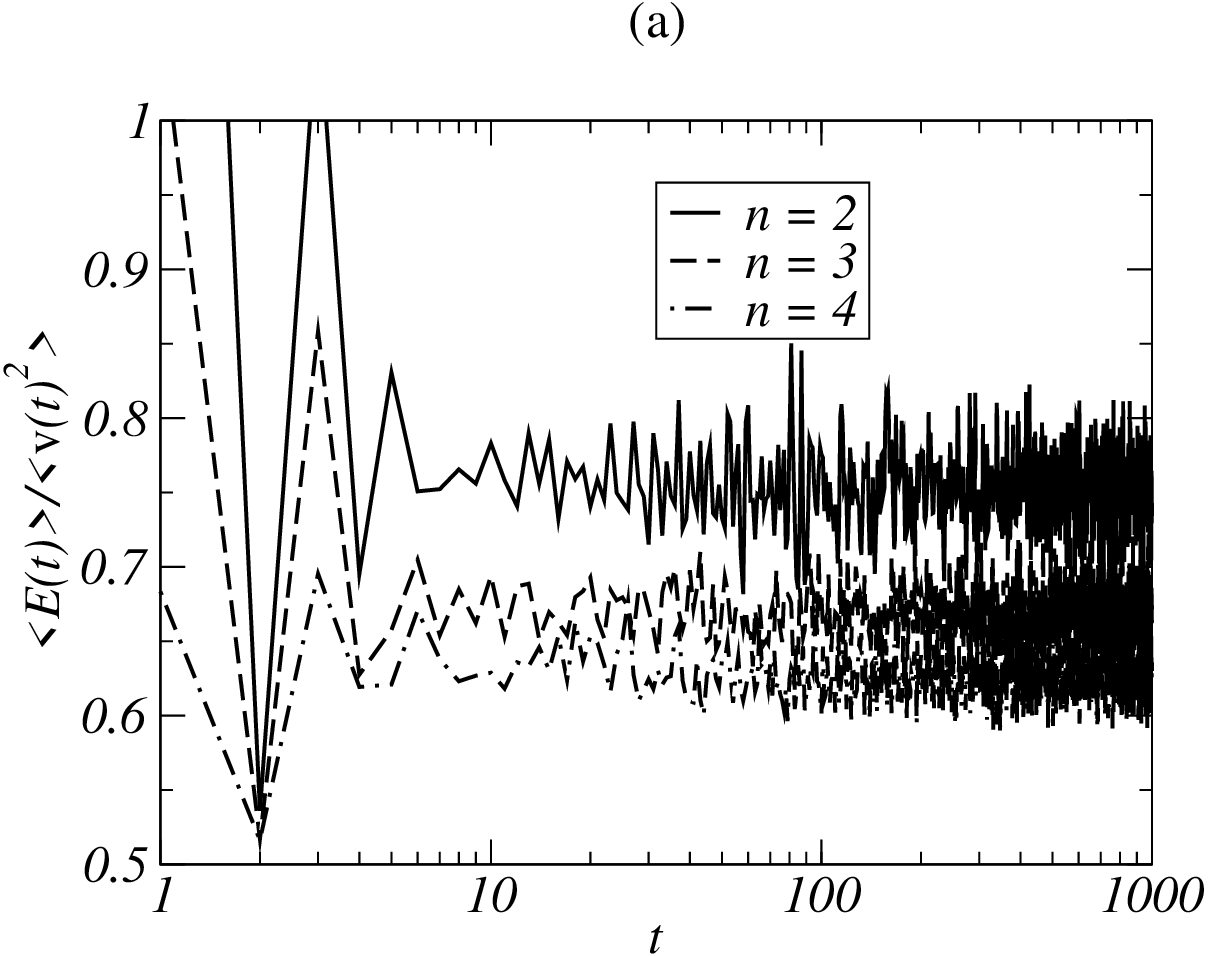}
\smallskip
\includegraphics*[width=0.75\columnwidth]{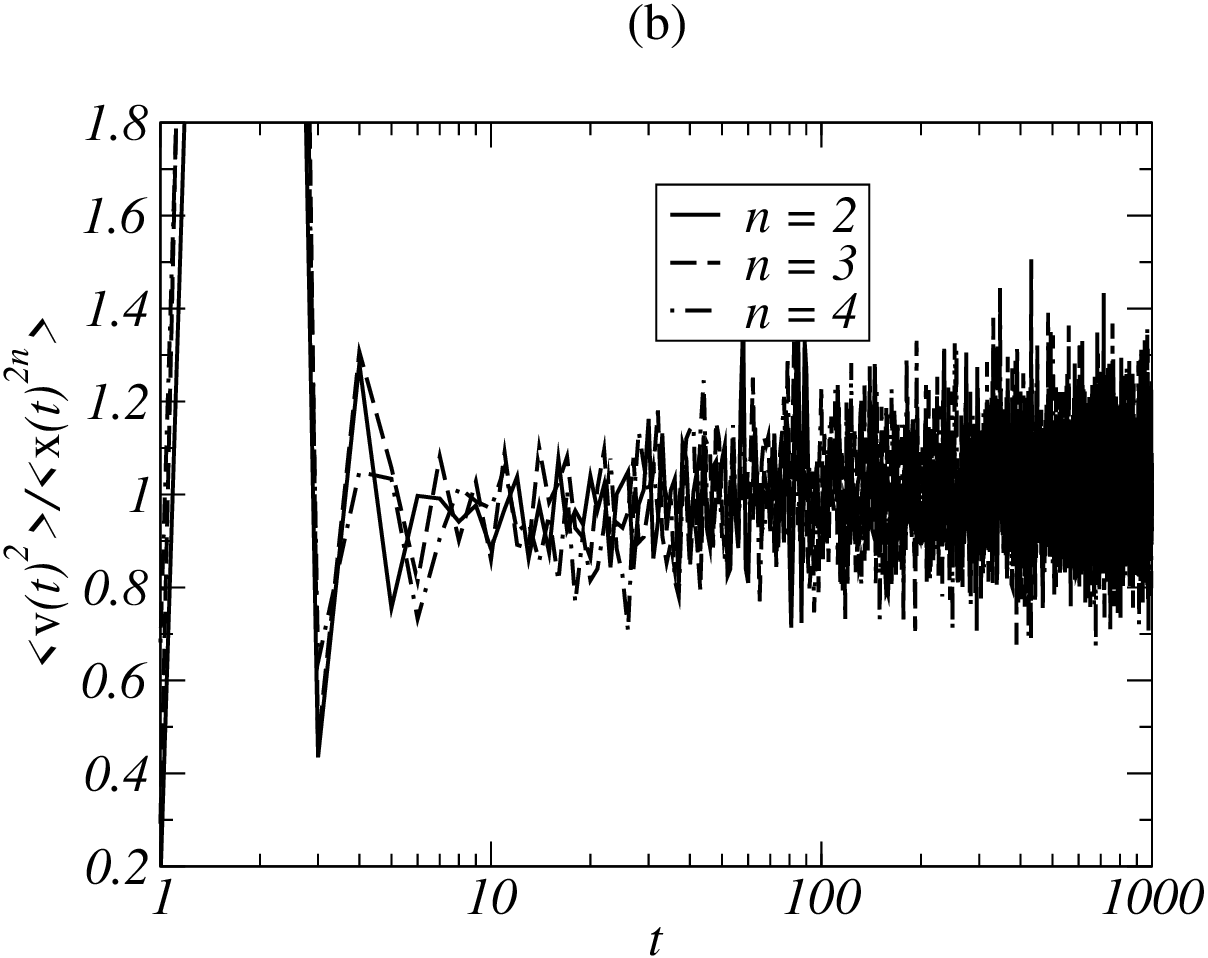}

\caption{Nonlinear oscillator with additive, colored noise:
Eqs.~(\ref{dyncolor})-(\ref{OU}) are integrated numerically 
for ${\mathcal D} = 1$ and $\tau = 1$, with a timestep 
$\delta t = 10^{-6}$ and averaged over $500$ realizations for
$n=2,3,4$. Fig.~(a): the first equipartition ratio 
$\langle E(t) \rangle / \langle v(t)^2 \rangle$ is close to
the theoretical value $\frac{n+1}{2n}$  given in  Eq.~(\ref{equipn2}).
Fig.~(b): the second equipartition ratio 
$\langle v(t)^2 \rangle / \langle x(t)^{2 n} \rangle$ is close to $1$.
}
\label{fig:colored:equip} 
\end{figure}

 We rewrite Eq.~(\ref{dyncolor})   in $(\Omega,\phi)$
  coordinates  and   obtain  as above 
[see Eqs.~(\ref{evoln1})  and (\ref{evoln})] 
   \begin{eqnarray}
     \dot \Omega  &=& n  \, {\mathcal S}_n'(\phi)  \, \eta(t)  
  \label{evolcol1}
   \,  ,   \\   
 \dot\phi  &=& 
\Big ( \frac{\Omega}{ (2n)^{\frac{1}{2n}}} \Big )^{\frac{n-1}{n}}
 - \frac{{\mathcal S}_n(\phi)}{\Omega}  \, \eta(t)  \, .\label{evolcol2}
   \end{eqnarray}
Again, the angle $\phi$ is a fast variable: we expect that 
the equipartition relations (\ref{equipn2}) and (\ref{equipnxv})
are also valid for co\-lored additive noise. 
This is indeed confirmed by numerical simulations
(See Figure~\ref{fig:colored:equip}).

The system (\ref{OU},\ref{evolcol1},\ref{evolcol2})
is equivalent to a Fokker-Planck equation for the P.D.F.
$P_t(\Omega, \phi, \eta)$. However, averaging this equation 
over $\phi$ does not lead to any conclusive result. 
The qualitative reason for this failure is as follows: when the period of 
the angular  variable becomes smaller than the coherence time $\tau$ 
of the noise, the noise itself is averaged out.

The goal of this section is to introduce alternative methods
to determine the scaling behavior of nonlinear oscillators
with colored additive noise, first qualitatively in 
Section~\ref{sec:colored:scaling} thanks to dimensional analysis,
then in a more rigorous manner in Section~\ref{sec:colored:selfconsistent}
using a self-consistent argument. Finally, we show in 
Section~\ref{sec:colored:crossover} that the crossover
between the behaviors typical of white noise and colored noise
can be embodied in  a single scaling function.

\subsection{Scaling analysis}
\label{sec:colored:scaling}

  The equations (\ref{dyncolor}) and (\ref{OU}) can be interpreted
 as a system of three coupled first-order stochastic differential equations
 in presence of  white noise. The corresponding Fokker-Planck equation
 for the P.D.F.   $P_t(x, v ,\eta)$ is given by
\begin{equation}
\frac{ \partial P}{\partial t} = - v\frac{\partial P}{\partial x}
 +  (x^{2n-1} - \eta)\frac{ \partial P}{\partial v}
 + \frac{1}{\tau} \frac{ \partial \eta P}{\partial \eta}
 +\frac{{\mathcal D}}{2 \tau^2}\frac{ \partial^2 P}{\partial \eta^2}\,  .
\label{FPcolor}
\end{equation}
 If we perform  a scaling  analysis in the spirit of \cite{yp}
 by  comparing terms in this equation two by two,
 we find that  a consistent balance is 
 found if $v^2 \sim x^{2n}$,  $x^{2n-1} \sim \eta$ and 
 $\eta^2 \sim {\mathcal D}t/{2 \, \tau^2}$. Eliminating $\eta$, we obtain
 the following scaling laws
\begin{eqnarray}
           E      &\sim&  \Bigg( \frac { {\mathcal D} t } { 2 \, \tau^2}\Bigg)
 ^{\frac{n}{(2n-1)}}  \, , \nonumber     \\
         \dot x  &\sim&  \Bigg( \frac { {\mathcal D} t } { 2 \, \tau^2}\Bigg)
 ^{\frac{n}{2(2n-1)} }  \,  , \nonumber    \\ 
          x      &\sim&  \Bigg( \frac { {\mathcal D} t } { 2 \, \tau^2}\Bigg)
^{\frac{1}{2(2n-1)}}  \, .
\label{scalingcolor}
\end{eqnarray}
A similar argument was used in \cite{philkir} in the case of colored
\emph{multiplicative} noise, and yielded different exponents (See Table 1).

\begin{figure}
\includegraphics*[width=0.75\columnwidth]{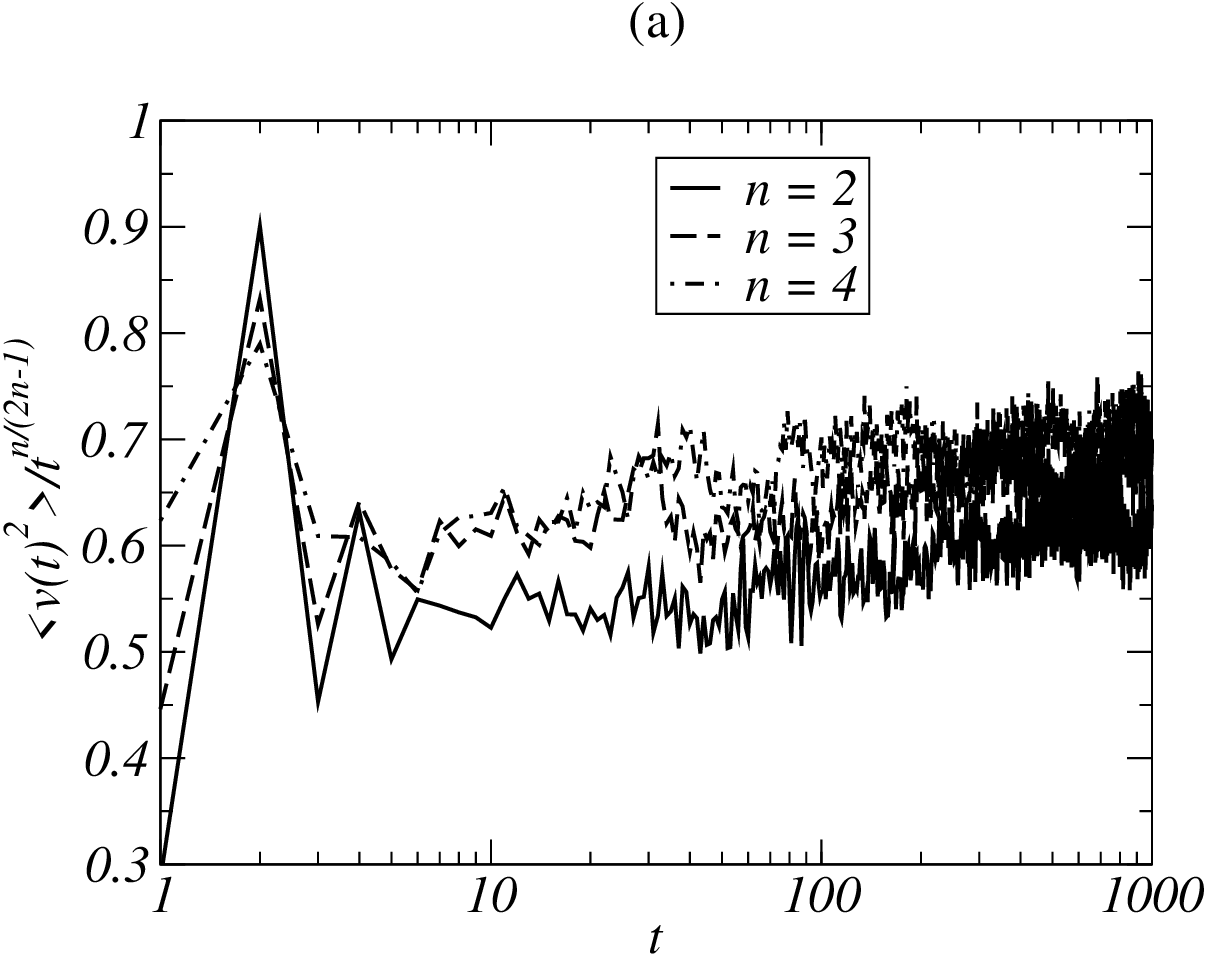}
\bigskip

\includegraphics*[width=0.75\columnwidth]{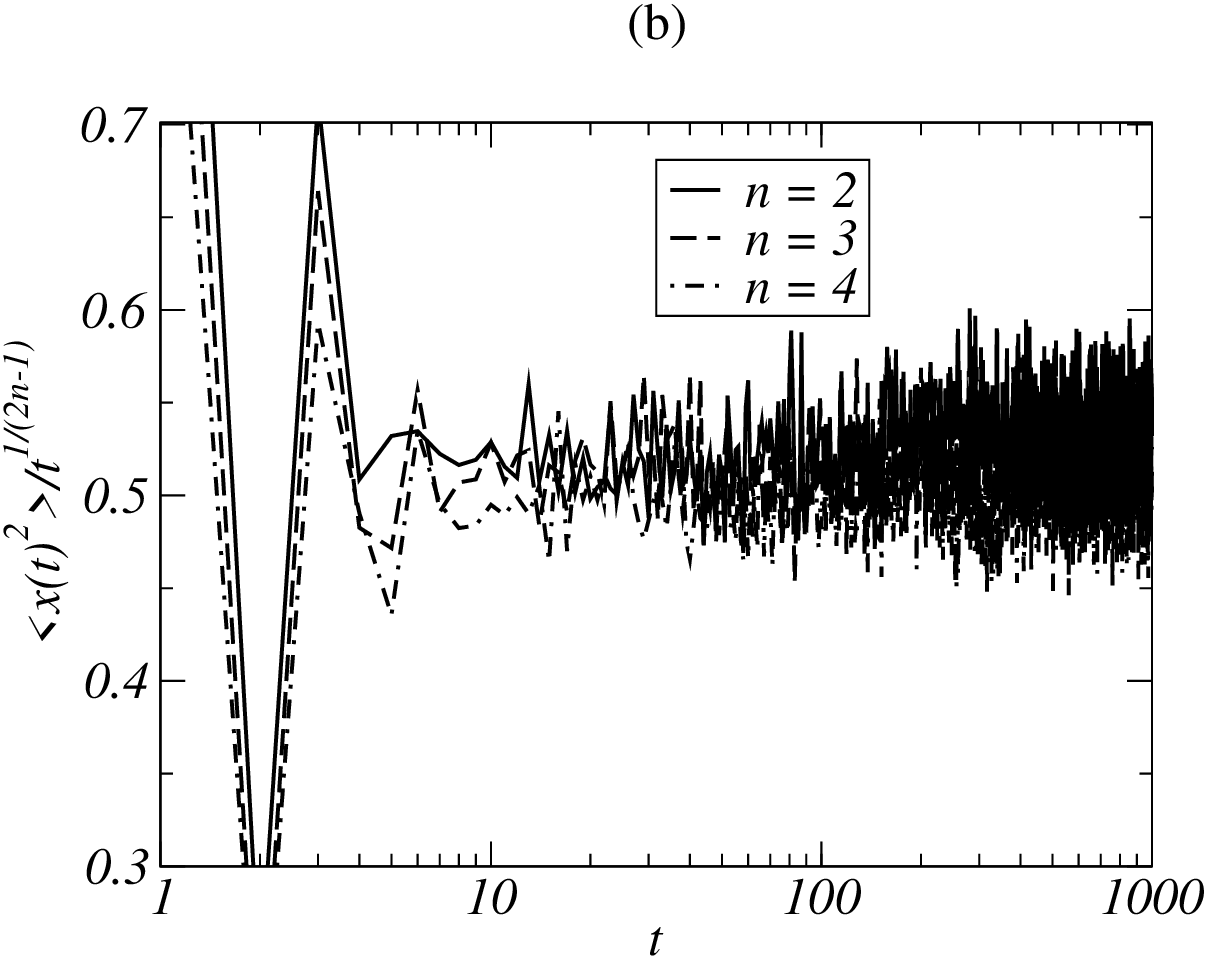}

\caption{Nonlinear oscillator with additive Ornstein-Uhlenbeck noise.
The ratios $\langle v^2 \rangle/t^{\frac{n}{2n-1}}$ (Fig.~(a))
and $\langle x^2 \rangle/t^{\frac{1}{2n -1}}$ (Fig.~(b)) are 
approximately constant at large time for $n = 2,3,4$.
Numerical data is obtained from the same simulations as in 
Fig.~\ref{fig:colored:equip}.
}
\label{fig:colored} 
\end{figure}

 Numerical simulations confirm the scalings predicted by
 Eqs.~(\ref{scalingcolor}) for general $n$ (See Figure~\ref{fig:colored}).
 Besides, these scalings are consistent with exact results 
found in the li\-near case $n=1$. 
 In the next section we present an 
 analytic derivation  of the exponents   appearing 
 in Eq.~(\ref{scalingcolor}).

\subsection{A self-consistent calculation}
\label{sec:colored:selfconsistent}

  We make the {\it a priori} Ansatz  that, in the long-time limit, 
 $\Omega$ grows  algebraically with time as
\begin{equation}
  \Omega \sim t^{\alpha}   \,, \label{autocst}
\end{equation} 
 and we will determine
 the scaling  exponent $\alpha$ from a self-consistent calculation.
 We deduce    from  Eq.~(\ref{evolcol2})  that $\phi~\sim~t^{\nu}$ where
\begin{equation}
       \nu =  \frac{n-1}{n} \alpha + 1  \, . \label{self1}
 \end{equation}
 Substituting  this scaling of $\phi$ in 
Eq.~(\ref{evolcol1}), we obtain   (hereafter we shall leave aside
 all  proportionality  constants)
 \begin{equation}
    \Omega \sim  \int_0^t \rm{d}z \; {\mathcal S}_n'(z^{\nu})  \, \eta(z)  \,.
 \label{eprcol}
\end{equation}

  We must now determine  the asymptotic  statistical behavior
  of this expression in the $ t \to \infty$ limit. In order
 to simplify the calculation, we replace the Ornstein-Uhlenbeck noise 
 by a  discrete dichotomous  noise, where time is discretized  in  steps of 
 duration $\tau$ (see \cite{chris} for a general discussion). 
 This approximation amounts to replacing 
  exponentially decaying  correlations, 
 by finite time correlations and therefore leaves
  the diffusion exponents unchanged.
 Equation (\ref{eprcol}) then  reduces to 
\begin{equation}
 \Omega  \sim  \sum_{k =0}^{t/\tau} \epsilon_k \int_{k\tau}^{(k+1)\tau}
    \rm{d}z \; {\mathcal S}_n'(z^{\nu})   \, ,
\label{eprdiscr}
  \end{equation}
 where the variable $\epsilon_k$ 
  takes the value $\pm \sqrt{  {\mathcal D}/(2 \tau)}$ randomly 
 during the time  interval $[k\tau \,  ,(k+1)\tau ]$; in other terms,
 \begin{equation}
   \langle  \epsilon_k \epsilon_l \rangle = 
  {  \frac {\mathcal D} {2 \tau} }\delta_{kl}   \, .
\label{bruitdiscr}
 \end{equation}
From Eqs.~(\ref{eprdiscr}) and (\ref{bruitdiscr}), we deduce that
 \begin{equation}
 \langle \Omega^2 \rangle \sim  \sum_{k =0}^{t/\tau}
  \Bigg(  \int_{k\tau}^{(k+1)\tau} 
   \rm{d}z \; {\mathcal S}_n'(z^{\nu})  \Bigg)^2 \, . \label{Om2}
\end{equation}
Integrating by parts  we obtain
 \begin{eqnarray}
  \int_{k\tau}^{(k+1)\tau} &\ & \rm{d}z \;  {\mathcal S}_n'(z^{\nu}) 
  =  \frac{ {\mathcal S}_n\Big( (k+1)^\nu\tau^\nu\Big) }{\nu 
     \Big((k+1)\tau\Big)^{\nu -1}  }\nonumber \\
&& - \frac{ {\mathcal S}_n\Big( k^\nu \tau^\nu \Big) }{\nu 
   \Big(k \tau \Big)^{\nu-1}  } \nonumber 
  + \frac{ \nu -1}{\nu} \int_{k\tau}^{(k+1)\tau} 
 \rm{d}z \; \frac {  {\mathcal S}_n(z^{\nu}) }{z^{\nu} }\label{evalk} \\
 \end{eqnarray} 
The integral term on the r.h.s. of (\ref{evalk})
is of order ${\mathcal O}( k^{-\nu})$.
Since ${\mathcal S}_n$ is a bounded function, the first two terms
are of order ${\mathcal O}( k^{-\nu+1})$, and will dominate
when $k \gg 1$ since $\nu > 1$.
We thus obtain
  \begin{equation}
  \langle \Omega^2 \rangle \sim  \sum_{k =1}^{t/\tau}
  \frac{1}{\Big(k \tau \Big)^{2\nu-2}  } \sim  t^{3 - 2\nu} \, .
 \label{evalOm2}
\end{equation}

Assuming that the variable $\Omega$ is not  multifractal,
we know   from the scaling hypothesis, Eq.~(\ref{autocst}),
that $\Omega^2  \sim  t^{2\alpha} $ 
   Comparing with Eq.~(\ref{evalOm2}),  we deduce that
 \begin{equation}
      2\alpha  =  3 - 2\nu   \, . \label{self2}
 \end{equation}
 Equations (\ref{self1}) and (\ref{self2})  provide the required 
 self-consistent condition and allow to calculate the exponents:
  \begin{equation}
   \alpha = \frac{n}{2(2n-1)} \,\,\, \hbox{ and }  \,\,\,
 \nu = \frac{5n - 3}{2(2n-1)} \, .
\end{equation}
 We can now  deduce from Eq.~(\ref{defOmega})
 and from the equipartition relations,
  Eqs.~(\ref{equipn2})  and  (\ref{equipnxv}),
  the scaling 
 behavior of the dynamical variables
 \begin{eqnarray}
  E  &\sim&  t^{ \frac{n}{2n-1} }  \,,   \nonumber \\ 
    v  &\sim&     t^{ \frac{n}{2(2n-1)}    }  \,,   \nonumber \\ 
   x  &\sim&  t^{ \frac{1}{2(2n-1)} } \, \,.
 \label{stochdynscal}
 \end{eqnarray}
 This result is in agreement with the 
 scaling analysis  of the previous section
  [see Equation (\ref{scalingcolor})].

\subsection{Crossover from  white  to colored  noise}
\label{sec:colored:crossover}

  We shall now match  the white  and the colored noise behaviors  
obtained in the long-time limit through
 a sca\-ling function  and  will describe  the crossover from the
 first regime to the second one. 
 In the system (\ref{dyncolor}) with Ornstein-Uh\-len\-beck noise, 
 one can define two  dimensionless variables: $t/\tau$ and
  ${\mathcal D}\tau^{\frac{3n-1}{n-1}}$.  Any  dimensionless
   quantity  can be written as a
    function of these two variables (or of any two
 independent combinations  of these two variables).
 For example,
  the mean value of the
 energy  can be written  in full generality,  as a function of time, as
 \begin{equation}
\langle E  \rangle  =   \frac{ {\mathcal D}t }{2} \; \Phi
 \Big({\mathcal D}\tau^{\frac{2n}{n-1}} t , \, 
 {\mathcal D}\tau^{\frac{3n-1}{n-1}} \Big) \,,
\label{scalAns}
\end{equation}
where the prefactor ${\mathcal D}t /2$ ensures that the equation
 is dimensionally correct and has been  chosen to match the
 white noise limit. The scaling function $\Phi$ allows to
 interpolate between  white and colored noise  and 
  exhibits the following asymptotic
 behavior:
 \begin{itemize}
\item The white noise limit   corresponds
 to  $\tau =  0$, therefore
 $\Phi(0,0) =  1$.
\item When $\tau$ is finite  (colored noise case)
 and $t \to \infty$, we deduce from
 Eq.~(\ref{scalingcolor}) that,  for $u \to \infty$
 and for $v$ finite,
 $\Phi(u,v) \to u^{\frac{1-n}{2n-1}} \phi(v)$. 
 The prefactor  $\phi(v)$  is  independent of time and  cannot
 be determined   by dimensional analysis. 
\end{itemize}
 The crossover between white noise and  colored noise is obtained
 when $u = {\mathcal D}\tau^{\frac{2n}{n-1}} t   \sim 1 \,,$ {\it i.e.},
 for  a typical time of the order of 
 \begin{equation}
  t_{\rm{cross}}\sim  \frac{ \tau^{-\frac{2n}{n-1}} }{ {\mathcal D}} \,.
\end{equation}
 At this  crossover  time  the 
  energy  is of  the order of 
 $ E \sim  {\mathcal D} t_{\rm{cross}}  \sim  \tau^{-\frac{2n}{n-1}}$.
  The period  of the deter\-mi\-nistic oscillator  for such an  energy 
  is  given by $T \sim E^{-\frac {n-1} {2n}} \sim \tau $
(see Eq.~(\ref{evolnphi})).
 Thus, the  crossover  occurs when the period of the deter\-mi\-nistic
 oscillator is of the order of the correlation time of the noise.
 This fact has   the following  intuitive explanation. When 
 $ t \ll  t_{\rm{cross}} $, the  period of  the deter\-mi\-nistic oscillator 
 is very large compared  to the correlation time of the noise: 
    the noise is totally uncorrelated over one period and
  can be viewed as a sequence  of  uncorrelated kicks 
 and thus  acts  as a  white noise.
 However, when   $ t \gg  t_{\rm{cross}} $,
 the  period of  the deterministic oscillator  is small 
 compared  to the correlation time of the noise:   the noise 
 is highly correlated over a period and can be viewed
  as an almost constant quantity  over  this time
 scale. Many periods must be added before 
  the effect of the noise   becomes perceptible:
  hence  the diffusion  slows down.

  We  remark  that  in Eq.~(\ref{scalAns})
  we   have  not used $ {t}/{\tau}$ as 
  the  dimensionless time   variable 
 which would have been  the most natural  choice.  In fact, 
 the white noise  limit $\tau \to 0$,   and  
 the long time   limit   $t \to \infty$ 
   with  finite $\tau$  are   indistinguishable  if  one uses 
 the variable $t /\tau$.
 For this reason, it may be incorrectly stated  that 
 `in the long  time limit,
 co\-lored  noise     appears as   white'. We have seen 
 in our problem  that  exactly the opposite is true: the effect of a non-zero
 correlation time becomes  relevant at long times.
 The limit $t /\tau \to \infty$ is singular and its value  depends on whether 
   the second  dimensionless variable
   $v = {\mathcal D}\tau^{\frac{3n-1}{n-1}}$  is equal to zero
 or is strictly positive. 
  By choosing  the variable
 $ u ={\mathcal D}\tau^{\frac{2n}{n-1}} t $  as the  dimensionless time 
 variable    in Eq.~(\ref{scalAns})   rather than  
    $ {t}/{\tau}$  we avoid  this difficulty  
   because  the white noise   and the colored noise cases  now   appear 
 as  different limits.

 \hfill\break

\begin{table}
  \centering 
  \begin{tabular}{||  c   ||  c   |  c  ||} 
  \hline 
                  &                &            \\  
  \hspace{15pt}           &     \hspace{15pt}     &   \hspace{15pt} 
         \\ 
 GAUSSIAN NOISE     &   WHITE   &   COLORED         \\ 
                 &           &                   \\ 
                   &                &            \\ 
\hline                  
                     &                &            \\ 
 & $x \sim   t^{\frac{1}{2n}}$
 & $x \sim   t^{\frac{1}{2(2n-1)}}$   \\ 
        &                &            \\ 
 ADDITIVE   &  $\dot x \sim   t^{\frac{1}{2}}  $    
     &    $\dot x \sim   t^{\frac{n}{2(2n-1)}}   $        \\ 
      &                &            \\ 
    &     $ E \sim \frac{{\cal D}}{2} t    $      
   &       $   E \sim  t^{\frac{n}{(2n-1)}}  $                \\ 
                  &                &            \\ 
 \hline            
                 &                &         \\  
          &   $  x \sim  t^{\frac{1}{2(n-1)}}           $     
       
     &      $  x \sim   t^{\frac{1}{4(n-1)}}   $               \\ 
            &                &         \\  
 MULTIPLICATIVE   &  $\dot x \sim    t^{\frac{n}{2(n-1)}}   $  
  &    $ \dot x \sim     t^{\frac{n}{4(n-1)}}   $            \\ 
                   &                &         \\  
                   &   $ E \sim  t^{\frac{n}{(n-1)}} $    
        &      $   E \sim   t^{\frac{n}{2(n-1)}} $           \\ 
     &                &         \\  
\hline 
\end{tabular}
\caption{Anomalous diffusion exponents for an undamped noisy nonlinear 
oscillator.}   
\end{table}

\section{Conclusion}

      A Hamiltonian nonlinear oscillator submitted to random internal noise
 (additive noise) exhibits anomalous diffusion phenomena.
 We have calculated analytically the associated scaling exponents
 in the case of white noise and  have obtained,  in the long time limit,
 an explicit expression for  the  probability
 distribution function   in phase spa\-ce. 
  The energy of such an oscillator   grows linearly
  with time irrespective of  the form of the confining potential.
 It is remarkable that  even the rate of  energy 
 growth does not depend on the stiffness  of the  potential and is 
 simply proportional to the amplitude  of the noise.  
Our results  also describe   the behavior of  an oscillator
 with small damping rate $\gamma$ for times  less than the dissipation time
 $\gamma^{-1}$. We have shown  that the   probability  distribution function 
of the energy of the undamped oscillator
 can be matched with   the  canonical Boltzmann-Gibbs  distribution   
 when $ t \sim \gamma^{-1}$.

 In the case of colored noise, the scaling exponents are modified.
 Their values have been determined by  a self-consistent
 calculation  and  the predictions of dimensional analysis
 have been confirmed. 
 The diffusion exponents are reduced because the coherence
 of the noise over a period of the system makes
 the energy transfer  less  efficient. The growth of energy
 is slower than linear and in the limit of  a very stiff potential, the energy
 grows only  as the square-root of time.

 In Table 1, we   summarize  and compare  the results   that we have derived 
 for additive noise  and   multiplicative noise, in this work and in
 \cite{philkir},   respectively.  In the case of white noise,
  precise results are available:  thanks to the averaging technique,
 the asymptotic  probability  distribution functions  are known.
 For colored noise, only the anomalous
  diffusion exponents have  so far  been
  calculated. We notice that  
 in all    cases  the exponent for the velocity is   half
  the exponent for the  energy and $n$ times the exponent for the
 amplitude {\it i.e.}, the ratios between corresponding
 exponents are independent of the problem considered.
 This  is the consequence of the generalized equipartition
 relations, which are independent of the nature of the noise 
  (in fact, equipartition  also provides  universal relations
 between  the generalized diffusion
 constants  appearing  as  prefactors 
  in  the  scaling  laws). Besides,   it was argued  in 
  \cite{philkir} that,  in the long time limit,  the 
   multiplicative noise always dominates over  the  additive noise
 and indeed we observe that
  the diffusion exponents for the  multiplicative noise
 are always  larger   than those for  the  additive noise. 

 Although  we have been able here  to derive   the exponents
 for   additive  colored noise, the generalized diffusion  constants,
 as  well as  the long time asymptotics 
  of the probability  distribution function 
  were not  found  by our approach and 
 their calculation  remains      an open problem.
  Besides,  the  precise  calculation of the  scaling function
 that describes the   crossover   between white and colored noise
  is certainly a challenging problem.

\begin{acknowledgement}
 We would like to thank Jean-Marc Luck for his constant interest in our work,
as well as Michel Bauer, Peter Holdsworth and Kunihiko Kaneko 
for useful comments. 
\end{acknowledgement}

\end{document}